\documentstyle[aps,twoside]{revtex}
\newcommand{\C}{{\if mm {{\rm C}\mkern -15mu{\phantom{\rm t}\vrule}}
\mkern +10mu \else \leavemode \hbox{I}\kern -.17em \hbox{C} \fi}}
\newcommand {\bj}{\mbox{\bf j}} 
\newcommand {\br}{\mbox{\bf r}}

\newcommand {\bx}{\mbox{\bf x}}

\newcommand {\bA}{\mbox{\bf A}}

\newcommand {\bp}{\mbox{\bf p}}
\newcommand {\p}{\partial}

\begin{document}
\title{Quadrupole Contribution in Semiclassical Radiation Theory}
\author{Bijan~Saha \\
{\small \it Laboratory of Information Technologies}\\
{\small \it Joint Institute for Nuclear Research}\\
{\small \it 141980, Dubna, Moscow reg., Russia\\
e-mail: saha@thsun1.jinr.ru, bijan@cv.jinr.ru}}
\maketitle
\begin{abstract}
Within the frame-work of semiclassical theory two-level approximation
in atomic system has been considered. Model proposed by M.D. Crisp and
E.T. Jaynes has been modified. It has been shown that the time-dependent
frequency shift depends on the higher multipole moments, retained in the 
Taylor expansion of electromagnetic field.
\end{abstract}
\vskip 3mm
\noindent
{\bf PACS} number(s): 03.65.S - semiclassical theories: quantum mechanics

\vskip 3mm
\noindent
\vskip 5mm
\noindent
Issues, related to the problem of interaction of photon  and 
micro-particle, in their full length are beyond the scope of 
quantum mechanics. They cannot be considered without invoking 
additional principles concerning the laws of occurrence and 
disappearance of electromagnetic field. 
According to quantum mechanics atom should remain in excited state 
for long in absence of external field, whereas experiment shows that
atom transforms into normal state emitting photon. This contradiction can
be explained if we take into account the fact that the moving electron
creates electromagnetic field which acts on the electron. 
Several authors tried to consider this reverse action of field on electron
several ways. One of these methods was proposed by Jaynes and Cummings
\cite{jaynes1} in 1963 that was further developed by Jaynes and Co and 
many others~\cite{crisp,jaynes2,berman,salmon,blaive,bosanac}. 
  
In classical electrodynamics the radiative process are calculated from 
self-energy of the electron in external fields. In contrast, in quantum 
electrodynamics, the self-energy is first thrown away and one begins
with bare particles; then the self-energy is put back in {\it photon by 
photon}, hence the use of perturbation theory. Recently, Barut and 
coauthors developed a quantum electrodynamics based on 
self-energy~\cite{barut1,barut2}.

Authors of the papers mentioned previously mainly confined their study 
within electric dipole moment. Here we make an attempt to enlarge this 
study taking into account the moments of higher order, particularly 
electric quadrupole moment. 

Let us consider a nonrelativistic, spinless particle in external 
magnetic field. It can be described by the Hamiltonian 
\begin{equation} 
\hat{H} = \frac{1}{2m} \bigl[\hat {\bp} 
- \frac{e}{c} \bA\bigr]^2 - \frac{e^2}{r} 
\end{equation} 
Varying this Hamiltonian with respect to $\bA$ and using the continuity
equation $ \frac{\p \rho}{\p t} + {\rm div} \bj = 0$ one finds 
\begin{eqnarray}
\bj & = & \frac{ie \hbar}{2m} \{ \Psi \nabla \Psi^{*} - 
\Psi^{*} \nabla \Psi \} - \frac{e^2}{mc}\bA \Psi \Psi^{*}\\
\rho & = & e\Psi^{*} \Psi 
\end{eqnarray}
Taking the field to be weak one we further neglect the diamagnetic term  
in the Hamiltonian and current density. 
Now, any state of atomic system may be expressed as
\begin{equation}
\Psi (\br, t) = \sum_{\alpha} a_\alpha (t) \psi_\alpha (\br)
\label{Psi}
\end{equation}
where $\psi (\br)$ is the eigen functions of 
$\hat{H}_0 = -(\hbar^2/2m) \nabla^2 - e^2/r$, i.e., 
$$\hat{H}_0 \psi_\alpha (\br) = E_\alpha \psi_\alpha (\br)$$ 
Putting (4) into (3) we obtain
\begin{equation}
\bj (t, \br) = \frac{e\hbar}{2mi}\sum_{\alpha, \beta} 
\bigl[\rho_{\alpha \beta} \psi_\beta^* \nabla \psi_\alpha -  
\rho_{\beta \alpha} \psi_\beta \nabla \psi_\alpha^*\bigr],
\end{equation} 
where 
\begin{equation}
\rho_{\beta\alpha} (t) = a_\alpha (t) a_\beta^* (t) =
\rho_{\alpha\beta}(t)^* 
\label{dm}
\end{equation}
is the $\beta\alpha$ density matrix element of the atom in the 
Schroediger picture that evolves according to
\begin{equation}
i \hbar \dot {\rho}_{\alpha\beta} (t) = \sum_{\gamma} 
[{\hat H}_{\alpha\gamma} \rho_{\gamma\beta} 
- \rho_{\alpha\gamma}{\hat H}_{\gamma\beta}]
\label{eedm}
\end{equation}
Since the magnetic field obeys the Maxwell equations, for $\bA$ in 
Coulomb gauge (${\rm div} \bA = 0$) we can write 
\begin{equation} 
\nabla^2 \bA - \frac{1}{c^2} 
\frac{\partial^2 \bA}{\partial t^2} = -\frac{4\pi}{c}\bj^{\bot} 
\end{equation}
Here $\bj^{\bot}$ is the transverse current density and defines as
$$ \bj^{\bot} = \frac{1}{4\pi}\nabla \times \nabla \times \int
\frac{\bj (t, \bx')}{|\bx -\bx'|} d^3 \bx' $$
Further we denote $\bj^{\bot} = \bj$. 
The solution to the Maxwell equation can be written as
\begin{equation}
\bA (\bx,t) = \frac{1}{c} \int \frac{\bj (\bx',t - |\bx -\bx'|/c)}
{|\bx - \bx'|} d^3 \bx'.
\end{equation}
Taylor expanding the expression for $\bj$ one gets
\begin{eqnarray}
\bA (\bx, t) \approx \int \frac{j (\bx', t)}
{|\bx -\bx'|} d^3 \bx'
- \frac{1}{c}\int \frac{\p j (\bx', t)}
{\p t}d^3 \bx'
+ \frac{1}{2c^2}\int \frac{\p^2 j (\bx', t)}
{\p t^2} |\bx -\bx'| d^3 \bx'+ \cdots
\end{eqnarray}
Further expanding $|\bx - \bx'|$ for $x' << x$ one finds
\begin{eqnarray}
\bA (\bx, t) &\approx& \int \frac{j (\bx', t)}
{|\bx -\bx'|} d^3 \bx'
- \frac{1}{c}\int \frac{\partial j (\bx', t)}
{\p t}d^3 \bx'
+ \frac{x}{2c^2}\int \frac{\p^2 j (\bx', t)}
{\p t^2} d^3 \bx' -
\frac{x}{2c^3 r} \int\ddot{j} x' d^3 \bx' - \cdots
\end{eqnarray}
Putting $\Psi = \sum_{\alpha}^{}a_\alpha (t) \psi_\alpha (\bx, t)$, 
where $\psi$: $H_0 \psi_\alpha = E_\alpha \psi_\alpha$ into the equation 
above and retaining the electric dipole and quadrupole  moments we find
\begin{eqnarray}
\bA (\bx, t) &\approx& \sum_{\alpha\beta} \rho_{\alpha\beta}(t) 
\Bigl[\frac{-i e\hbar}{2\pi^2 mc} \int\limits_{0}^{\infty} dk \int 
d\Omega (\beta|e^{-ik\cdot x'}
\nabla | \alpha)_{\bot} e^{ik\cdot x} \nonumber \\
&+& \bigl( \frac{2}{3 c^2} \Omega_{\alpha\beta}^{} + \frac{i r}{3 c^3} 
\Omega_{\alpha\beta}^{3}\bigr) D_{\alpha\beta}^{(1)} -
\frac{i x_\alpha}{2 c^3 r} \Omega_{\alpha\beta}^{3} D_{\alpha\beta}^
{(2)}\Bigr]
+ \bA_0 (\bx, t)
\end{eqnarray}
where the transition frequencies and the electric dipole and
quadrupole moments are defined, respectively, as 
\begin{mathletters}
\begin{eqnarray}
\Omega_{\alpha\beta} &=& (E_\alpha - E_\beta)/\hbar, \label{tf}\\
{\bf D}_{\alpha\beta} &=& \int \psi_\alpha e {\bf x} \psi_{\beta}^{*} d\bx,
\quad {\rm or\,\, in\,\, components} \quad
D_{\alpha\beta}^{(i)} = \int \psi_\alpha e x^{i} \psi_{\beta}^{*} d\bx
\label{dipole} \\
Q_{\alpha\beta}^{(ij)} &=& \int  
\psi_\alpha e r^{ij} \psi_{\beta}^{*} d\bx,
\qquad  
r^{ij} = \frac{1}{2}(x^i x^j - \frac{1}{3}r^2
\delta^{ij}), \quad r = |\bx|, \label{quadrupole}
\end{eqnarray}
\end{mathletters}
Here $\bA_0$ is an externally applied field.
Putting the expression for $\bA$ into (\ref{eedm}), for density matrix we 
find
\begin{eqnarray}
\label{dmn}
\dot\rho_{\alpha\beta} &=& - i \Omega_{\alpha\beta} \rho_{\alpha\beta}
- i \sum_{\kappa} (\Gamma_{\alpha\kappa} - \Gamma_{\kappa\beta})
\rho_{\kappa\kappa}\rho_{\alpha\beta} \nonumber \\
&-& \sum_{\kappa}\bigl[
\frac{1}{2}({\cal A}_{\alpha\kappa} + {\cal A}_{\beta\kappa})
- ({\cal B}_{\alpha\kappa} + {\cal B}_{\beta\kappa})
+ ({\cal C}_{\alpha\kappa} + {\cal C}_{\beta\kappa})\bigr]
\rho_{\kappa\kappa} \rho_{\alpha\beta} \\
&-& \frac{\bA_0 (0,t)}{\hbar c}\sum_{\kappa}
\bigl[\Omega_{\alpha\kappa}{\bf D}_{\alpha\kappa} \rho_{\kappa\beta} 
- \Omega_{\kappa\beta}{\bf D}_{\kappa\beta} \rho_{\alpha\kappa}\bigr],
\nonumber
\end{eqnarray}
where we define
\begin{mathletters}
\begin{eqnarray}
\Gamma_{\alpha\beta} &\equiv& -\frac{e^2 \hbar}{2\pi^2 m^2 c^2}\int\limits_{0}^{\infty}
\int d\Omega (\alpha|e^{ik\cdot x'}|\beta)_{\bot} 
(\beta|e^{-ik\cdot x}|\alpha)_{\bot} =
\Gamma_{\beta\alpha}, \label{Gamma} \\
{\cal A}_{\alpha\beta}&\equiv& \frac{4}{3}({\bf D}_{\alpha\beta} 
{\bf D}_{\beta\alpha}/\hbar c^3)
\Omega_{\alpha\beta}^{3} = -{\cal A}_{\beta\alpha}, \quad
{\rm Einstein\,\, coefficient} \label{Ein} \\
{\cal B}_{\alpha\beta}&\equiv& ({\bf D}_{\alpha\beta} \Delta_{\alpha\beta}
/\hbar c^4)\Omega_{\alpha\beta}^{3} \equiv, -{\cal B}_{\beta\alpha},\qquad 
\Delta_{\alpha\beta} = \int r \bar{J}_{\alpha\beta} (\bx) d\bx, \label{cB} \\
{\cal C}_{\alpha\beta}&\equiv& (Q_{\alpha\beta}^{ij} 
\delta_{\alpha\beta}^{k}/\hbar c^4)
\Omega_{\alpha\beta}^{3} \equiv -{\cal C}_{\beta\alpha},
\qquad \delta_{\alpha\beta}^{k} = 
\int \frac{x^k}{r} \bar{J}_{\alpha\beta} (\bx) d\bx.  \label{cC}
\end{eqnarray}
\end{mathletters}
Here we denote $\bar{J}_{\alpha\beta} = (e\hbar/2mi)
\bigl[\psi_\beta^* \nabla\psi_\alpha - \psi_\beta\nabla\psi_\alpha^*\bigr]$.

The equation (\ref{dmn}) can be written in the following way where the 
repeating index denotes summation
\begin{eqnarray}
\label{dm1}
\dot\rho_{\alpha\beta} &=& - i \Omega_{\alpha\beta} \rho_{\gamma\tau}
M_{\alpha\beta\gamma\tau}
- i (\Gamma_{\alpha\kappa} - \Gamma_{\kappa\beta})
\rho_{\kappa\kappa}\rho_{\gamma\tau}M_{\alpha\beta\gamma\tau} \nonumber \\
&-& \bigl[\frac{1}{2}({\cal A}_{\alpha\kappa} + {\cal A}_{\beta\kappa})
- ({\cal B}_{\alpha\kappa} + {\cal B}_{\beta\kappa})
+ ({\cal C}_{\alpha\kappa} + {\cal C}_{\beta\kappa})\bigr]
\rho_{\kappa\kappa} \rho_{\gamma\tau}M_{\alpha\beta\gamma\tau} \\
&-& \frac{A_0 (0,t)}{\hbar c}
\bigl[\Omega_{\alpha\kappa}D_{\gamma\kappa}^{(1)} \rho_{\kappa\tau} 
- \Omega_{\kappa\beta}D_{\kappa\tau}^{(1)} \rho_{\gamma\kappa}\bigr]
M_{\alpha\beta\gamma\tau},
\nonumber
\end{eqnarray}
where $M_{\alpha\beta\gamma\tau} = \delta_{\alpha\gamma} \delta_{\beta\tau}.$

As one sees from (\ref{dmn}) or (\ref{dm1}), the off-diagonal density matrix 
elements oscillate at frequencies 
$\Omega_{\alpha\beta} + \delta \Omega_{\alpha\beta}(t)$, where the 
time-dependent frequency-shift is
\begin{equation}
\delta \Omega_{\alpha\beta}(t) = - \sum_{\kappa}
 (\Gamma_{\alpha\kappa} - \Gamma_{\kappa\beta}) \rho_{\kappa\kappa} (t)
\label{fs}
\end{equation}
Now the expectation of dipole moment of the atom
\begin{eqnarray}
<{\bf \mu}> = \int \Psi^{*}({\bf x}, t) e {\bf x} \Psi({\bf x}, t) d\bx
\nonumber
\end{eqnarray}
in account of (\ref{Psi}) can be written as
\begin{eqnarray}
<{\bf \mu}> = \sum_{\alpha\beta} {\bf D}_{\alpha\beta} \rho_{\beta\alpha}(t).
\nonumber
\end{eqnarray}
Thus we see that the off-diagonal matrix elements are directly connected
with the expectation of dipole moment.

In what follows we take into account only two of these levels. We choose
the zero from which we measure the energies to be midway between the
two active levels, so that
\begin{equation}
E_2 = - E_1
\label{mid}
\end{equation}
The equation (\ref{dm1}) can then be written as
\begin{mathletters}
\label{comp}
\begin{eqnarray}
\dot{\rho}_{11} &=& - 2 q \rho_{11}\rho_{22}\\
\dot{\rho}_{22} &=&  2 q \rho_{11}\rho_{22}\\
\dot{\rho}_{12} &=& -i \bigl[\Omega_{12} + 
\Gamma_{11}\rho_{11} -\Gamma_{22}\rho_{22}
-\Gamma_{12} (\rho_{11} -\rho_{22})\bigr] \rho_{12}
+  q (\rho_{11} - \rho_{22}) \rho_{12}\\
\dot{\rho}_{21} &=& -i \bigl[\Omega_{21} - \Gamma_{11}\rho_{11} 
+\Gamma_{22}\rho_{22}
+\Gamma_{12} (\rho_{11} -\rho_{22})\bigr] \rho_{21}
+  q (\rho_{11} - \rho_{22}) \rho_{21}
\end{eqnarray}
\end{mathletters}
where we denote $2 q = {\cal A}_{12} - 2 {\cal B}_{12}+ 2 {\cal C}_{12}$.

Let us now rewrite $\rho_{\alpha\beta}$ in the form~\cite{fano1,fano2}
\begin{equation}
\rho_{\alpha\beta} = \frac{1}{2} \bigl(\delta_{\alpha\beta} 
+ P_{j} \sigma_{\alpha\beta}^{j}\bigr)
\label{fano}
\end{equation}
where $\sigma^{j}$ are the Pauli matrices and 
${\bf P} = (P_x, P_y, P_z)$ is a unit vector of three-dimensional 
Poincar$\acute e$ representation.
From ~(\ref{fano}) follow:
\begin{eqnarray}
\label{fano1}
\rho_{11} &=& \frac{1}{2} \bigl(1 + P_z), \quad 
\rho_{12} = \frac{1}{2} \bigl(P_x - i P_y), \nonumber\\ \\
\rho_{22} &=& \frac{1}{2} \bigl(1 - P_z), \quad
\rho_{21} = \frac{1}{2} \bigl(P_x + i P_y) \nonumber
\end{eqnarray}
or equivalently,
\begin{equation}
\rho_{11} + \rho_{22} = 1,\quad
\rho_{11} - \rho_{22} = P_z, \quad \rho_{12} + \rho_{21} = P_x, \quad
\rho_{12} - \rho_{21} = - i P_y
\label{fano2}
\end{equation}
In account of (\ref{fano1}) and (\ref{fano2}) from (\ref{comp}) we find the 
following system of equations
\begin{mathletters}
\label{final}
\begin{eqnarray}
\dot{P}_x &=& q P_z P_x + (\Omega_{12} + \tau + \lambda P_z) P_y \\
\dot{P}_y &=& q P_z P_y - (\Omega_{12} + \tau + \lambda P_z) P_x \\
\dot{P}_z &=& q (P_z^2 - 1)
\end{eqnarray}
\end{mathletters}
where we denote $\tau = (\Gamma_{11} - \Gamma_{22})/2$ and
$\lambda = (\Gamma_{22} + \Gamma_{11})/2 - \Gamma_{12}$.
The solutions to the system of equations (\ref{final}) read
\begin{mathletters}
\label{P}
\begin{eqnarray}
P_x &=& {\rm cos}\,[\Omega_{12} (t - t_0) + \tau (t - t_0) + 
(\lambda /q) {\rm ln\, cosh}\, q(t - t_0)]\, {\rm sech}\, q(t - t_0)\\
P_y &=& {\rm sin}\,[\Omega_{12} (t - t_0) + \tau (t - t_0) + 
(\lambda /q) {\rm ln\, cosh}\, q(t - t_0)]\, {\rm sech}\, q(t - t_0)\\
P_z &=& -{\rm tanh}\, q(t - t_0)
\end{eqnarray}
\end{mathletters}
Rewriting (\ref{P}) in terms of $\rho$ we find
\begin{mathletters}
\label{}
\begin{eqnarray}
\rho_{11}&=&1/\Bigl[{\rm exp}\,[2q(t - t_0)] +1\Bigr],\\
\rho_{22}&=&1/\Bigl[{\rm exp}\,[-2q(t - t_0)] +1\Bigr],\\
\rho_{12}&=&\Bigl[{\rm exp}\,\Bigl(-i [\Omega_{12} (t - t_0) + 
\tau (t - t_0) + (\lambda /q) 
{\rm ln\, cosh}\,q(t - t_0)]\Bigr)\Bigr]\,{\rm sech}\, q(t - t_0),\\      
\rho_{21}&=&\Bigl[{\rm exp}\,\Bigl(i [\Omega_{12} (t - t_0) + 
\tau (t - t_0) + (\lambda /q) 
{\rm ln\, cosh}\,q(t - t_0)]\Bigr)\Bigr]\,{\rm sech}\, q(t - t_0).      
\end{eqnarray}
\end{mathletters}
For the expectation value of the energy in account of (\ref{mid}) we find
\begin{eqnarray}
<H_0> &=& E_1 \rho_{11} (t) + E_2 \rho_{22} (t) =
     - \frac{\hbar}{2} \Omega_{21} (\rho_{22} - \rho_{11}) \nonumber \\
 &=& - \frac{\hbar}{2} \Omega_{21} {\rm tanh} [q(t - t_0)].
\label{H0}
\end{eqnarray}
where as, for the expectation of the dipole moment we obtain
\begin{eqnarray}
<{\bf \mu}> &=& {\bf D}_{21} (\rho_{12} + \rho_{21})=
{\bf D}_{21} P_x \nonumber\\
&=& {\bf D}_{21}\,{\rm sech}\, q(t - t_0)\,
{\rm cos}\,[\Omega_{21} t + \vartheta(t)],
\end{eqnarray}
where we define
\begin{eqnarray}
\vartheta (t) = \vartheta_0 -\tau t
- (\lambda/q) {\rm ln\, cosh} q(t - t_0), \quad
\vartheta_0 = [(\Gamma_{11} - \Gamma_{22})/2 - \Omega_{21}] t_0
\label{theta}
\end{eqnarray}
and corresponds to a time-dependent frequency shift
\begin{equation}
\delta \Omega_{21} (t) = d\vartheta/d t = -\tau - \lambda {\rm tanh} q(t-t_0)
\label{fre-s}
\end{equation}
Comparing (\ref{fre-s}) with those obtained in ~\cite{crisp} one finds
the additional frequency shift as 
\begin{equation}
\Delta (\delta \Omega_{21} (t)) = \lambda 
\frac{{\rm tanh} [({\cal C}_{21} - {\cal B}_{21})(t - t_0)] {\rm sech}^2
[{\cal A}_{21}(t - t_0)/2]}
{1 +{\rm tanh}[{\cal A}_{21}(t - t_0)/2]{\rm tanh}[({\cal C}_{21} - {\cal B}_{21})(t - t_0)] }
\label{pq}
\end{equation}
Thus we see that beside Einstein  $A$ coefficient, higher multipole
moments, in particular quadrupole one, contribute to the spontaneous 
decay of the atom from an exited state.

{\bf Acknowledgement:} Few years back Prof. M.A. Martsenyuk first drew my 
attention to this problem. Further I was also inspired by Prof. M.D. Crisp. 
Taking this opportunity, I'd like to thank both of them.

\end{document}